\documentclass[useAMS,usenatbib]{mnras}

\usepackage{times}
\usepackage{graphics,epsfig}
\usepackage{graphicx}
\usepackage{amsmath}
\usepackage{amssymb}
\usepackage{bm}
\usepackage{dcolumn}
\usepackage{epsfig}
\usepackage{subfig}
\usepackage{tabularx}
\usepackage[usenames,dvipsnames,svgnames,table]{xcolor}

\newcommand{\kms}{$\rm km\,s^{-1}$}

\newcommand{\ha}{\ensuremath{{\rm H}\alpha~}}

\newcommand{\arcs}{\ensuremath{^{\prime\prime}}}

\definecolor{bondiblue}{rgb}{0.0, 0.58, 0.71}

\definecolor{cerulean}{rgb}{0.0, 0.48, 0.65}

\definecolor{darkcerulean}{rgb}{0.03, 0.27, 0.49}

\title[A Molecular gas rich GRB host]
{A Molecular gas rich GRB host galaxy at the peak of cosmic star formation \\
}
\author[Arabsalmani et al.]
{M. Arabsalmani$^{1,2,3}$\thanks{E-mail: maryam.arabsalmani@cea.fr}, E. Le Floc'h$^{1,2}$,  H. Dannerbauer$^{4,5}$, C. Feruglio$^{6}$, E. Daddi$^{1,2}$, \\
\newauthor L. Ciesla$^{1,2}$, V. Charmandaris$^{7,8}$, J. Japelj$^9$, S. D. Vergani$^3$, P.-A. Duc$^{10,1,2}$, S. Basa$^{11}$,\\
\newauthor      F. Bournaud$^{1,2}$, D. Elbaz$^{1,2}$  \\\\
        $^1$ IRFU, CEA, Universit\'e Paris-Saclay, F-91191 Gif-sur-Yvette, France\\
        $^2$ Universit\'e Paris Diderot, AIM, Sorbonne Paris Cit\'e, CEA, CNRS, F-91191 Gif-sur-Yvette, France\\
        $^3$ GEPI, Observatoire de Paris, PSL Research University, CNRS, Place Jules Janssen, 92190 Meudon, France\\
        $^4$ Instituto de Astrofísica de Canarias (IAC), E-38205 La Laguna, Tenerife, Spain \\
        $^5$ Universidad de La Laguna, Dpto. Astrofísica, E-38206 La Laguna, Tenerife, Spain\\
        $^{6}$ INAF Osservatorio Astronomico di Trieste, Via G. Tiepolo 11 , I-34124 Trieste \\
        $^7$ Institute for Astronomy, Astrophysics, Space Applications \& Remote Sensing, National Observatory of Athens, GR-15236, Penteli, Greece\\
        $^8$ Department of Physics, University of Crete, GR-71003 Heraklion, Greece\\
        $^9$ Anton Pannekoek Institute for Astronomy, University of Amsterdam, Science Park 904, 1098 XH Amsterdam, The Netherlands\\
        $^{10}$ Universit\'e de Strasbourg, CNRS, Observatoire astronomique de Strasbourg, UMR 7550, F-67000 Strasbourg, France\\
        $^{11}$ LAM - Laboratoire d'Astrophysique de Marseille, 13388 Marseille Cedex 13, France
}

\begin{document}
\date{}

\pagerange{\pageref{firstpage}--\pageref{lastpage}} \pubyear{}

\maketitle

\label{firstpage}

\begin{abstract}
We report the detection of the CO(3-2)  emission line from the   host galaxy of  Gamma Ray Burst (GRB) 080207 at $z$ = 2.086. This is the first detection of molecular gas in emission from a GRB host galaxy beyond redshift 1. We find this galaxy to be rich in molecular gas  with a  mass of $1.1 \times 10^{11}\,\rm M_{\odot}$ assuming $\alpha_{\rm CO}=$ 4.36 $\rm M_{\odot}(\rm K\,km\,s^{-1}\,pc^2)^{-1}$. The molecular gas mass  fraction of the galaxy is $\sim$ 0.5,  typical of star forming galaxies (SFGs) with similar stellar masses and redshifts. With a $\rm SFR_{FIR}$ of 260 $\rm M_{\odot}\,yr^{-1}$, we measure a molecular-gas-depletion timescale of  0.43 Gyr, near the peak of the depletion timescale distribution of SFGs at similar redshifts. 
Our findings  are therefore  in contradiction  with the proposed molecular gas deficiency in GRB host galaxies. We argue that the reported molecular gas deficiency for  GRB hosts could be the  artifact of  improper comparisons or neglecting the effect of the typical low metallicities  of GRB hosts on  the  CO-to-molecular-gas conversion factor. We also compare the kinematics of the CO(3-2) emission line  to that  of the \ha emission line from the host galaxy. We find  the \ha emission to have contributions from  two separate components,  a narrow and a broad one.  The narrow component matches  the CO emission well in velocity space. The broad component, with  a FWHM  of $\sim$ 1100 \kms, is separated by  $+390$ \kms\, in velocity space from the  narrow component. We speculate this broad component to be associated with a powerful outflow in the host galaxy or in an interacting system.


\end{abstract}



\section{Introduction}
\label{sec:intro}
Long-duration Gamma Ray Bursts (GRBs) are believed to originate in massive stars and hence are beacons of star-forming galaxies 
\citep[e.g.][]{Sokolov01-2001A&A...372..438S, Lefloch03-2003A&A...400..499L, Fruchter06-2006Natur.441..463F}. 
The detectability of these extremely bright and dust-penetrating explosions is independent
of the brightness and dust content of their host galaxies. Hence they provide a unique method for sampling star-forming galaxies
throughout the Universe without a luminosity bias, a challenge  that significantly impacts even the deepest flux-limited galaxy surveys. 
GRB host galaxies  typically show very high  H{\sc i} column densities (N(H{\sc i}) $> 10^{21}\,\rm cm^{-2}$)  measured through 
the Lyman-$\alpha$ absorption line in GRBs spectra  \citep[e.g.,][]{Jakobsson06-2006A&A...460L..13J, Prochaska07-2007ApJ...666..267P, Fynbo09-2009ApJS..185..526F}.    
Contrarily, molecular gas in absorption has been detected in the spectra of only four GRBs \citep[][]{Prochaska09-2009ApJ...691L..27P, Kruhler13-2013A&A...557A..18K, Delia14-2014A&A...564A..38D, Friis15-2015MNRAS.451..167F}.  
Dissociation by GRB emission radiation is ruled out as the cause of the apparent low detection rate of  molecular gas in GRB afterglows  
given the large distances between GRB locations and  gas detected  in absorption \citep[larger than several hundreds of pc, see][]{Ledoux09-2009A&A...506..661L}. Studies of absorbing systems  in  lines of sight towards quasars  show that 
the detection of molecular gas 
is not coupled to the high N(H{\sc i}), but rather to high metallicities and  depletion factors 
which increases the formation rate of H$_2$ onto dust grains 
\citep[see][]{Noterdaeme08-2008A&A...481..327N}. 
This is consistent with the low detection rate of molecular lines in GRB afterglows  \citep[][]{Ledoux09-2009A&A...506..661L, Kruhler13-2013A&A...557A..18K} 
as GRB host galaxies typically have low metallicities \citep[e.g.][]{Savaglio09-2009ApJ...691..182S, Fynbo09-2009ApJS..185..526F}. 

Detection of molecular gas in emission  has been reported   for three GRB host galaxies at redshifts $z$ =  0.0087, 0.089, and 0.81  \citetext{\citealp{Hatsukade14-2014Natur.510..247H}; \citealp{Stanway15-2015ApJ...798L...7S}; \citealp{Michalowski16-2016A&A...595A..72M}; \citealp[see also][]{Perley17-2017MNRAS.465L..89P}}. All  three hosts are reported  to be deficient in molecular 
gas (with respect to  their stellar masses or star-formation rates), with  short molecular-gas-depletion times. This is proposed to be   characteristic of the  GRB population, 
suggesting formation of GRBs  toward the end of the star formation episodes of their host galaxies \citep[see][]{Stanway15-2015ApJ...798L...7S}. 
Intense star-formation  in the GRB environment \citep[][]{Hatsukade14-2014Natur.510..247H} as well as  formation of stars from atomic gas 
before its conversion to molecular gas \citep[][]{Michalowski16-2016A&A...595A..72M} are 
alternative hypotheses  proposed to explain the apparent tendency of GRB hosts to be deficient in molecular gas. 

In this paper we present the properties of molecular gas in the host galaxy of  GRB 080207 at $z$ = 2.086  
 through the observations performed   with the Plateau de Bure / NOrthern Extended Millimeter Array (NOEMA) at 112 GHz 
and also   with the Large Apex BOlometer CAmera (LABOCA) on the Atacama Pathfinder Experiment (APEX) at 870 $\mu$m \footnote{Based on observations made with ESO telescopes at the La Silla Paranal Observatory under programme ID 089.F-9304.}. 
The host galaxy of GRB 080207  is a massive, extremely red and dusty star-forming  galaxy 
\citep{Hunt11-2011ApJ...736L..36H, Rossi12-2012A&A...545A..77R}, making it a promising candidate for detection of molecular gas emission lines. 
The redshift of the galaxy being at the peak of the cosmic star-formation  makes it even more interesting for studies 
of molecular gas, the fuel of star formation. 
The observations and data reduction are described in  Section \ref{sec:obs}. Our findings and results are presented in Section \ref{sec:res}. 
In Section \ref{sec:dis} we discuss the reported molecular gas deficiency in GRB host galaxies, and finally  in Section \ref{sec:sum} we present a summary. 
Throughout this paper we use  a flat $\Lambda$CDM with H$_0$=69.6 \kms\,Mpc$^{-1}$ and $\Omega_m$=0.286. 

  
\section{Observations and data reduction}
\label{sec:obs}  

We targeted the  host galaxy of GRB 080207   for the CO(3-2) transition with Plateau de Bure / NOEMA  in C-configuration (project code w0c9) in May 2013 using a bandwidth of 2.3 GHz, centred  at the red-shifted CO(3-2) line frequency 
of  112.06\,GHz. 
The observations were performed with the dual polarization mode and with a total on-source time of 9.12 hours. Data were reduced and analyzed with the Grenoble Image and Line Data Analysis Software. 
After flagging and  calibration process, the calibrated  visibilities were mapped  to 
produce a spectral cube  assuming natural weighting, reaching a rms-noise of 0.7\,mJy/beam in line-free channels with 50\,MHz width and a  synthesized beam of 5.1\arcsec $\times$ 4.5\arcs. 
For the continuum of the galaxy we reached a  $3\sigma$ upper limit of 0.16 mJy/beam, without detecting the galaxy. 

We also observed this host  at 870 $\mu$m with the bolometer camera LABOCA \citep[][]{Siringo09-2009A&A...497..945S} on the APEX telescope (program ID 089.F-9304). The observations were carried out in April and August 2012 in service mode with typical pwv´s between 0.5-1.8 mm. 
We observed our target in point source photometry mode (where the target is placed on a reference pixel) for a total observing time of 4.2 hours. Data reduction was performed with the CRUSH software \citep[][]{Kovacs08-2008SPIE.7020E..1SK}, which led to a 
non-detection  with a $3\sigma$ upper limit flux of   9.0 mJy for the galaxy at 870 $\mu$m.

\begin{figure}
\begin{center} \hskip -5mm
\psfig{file=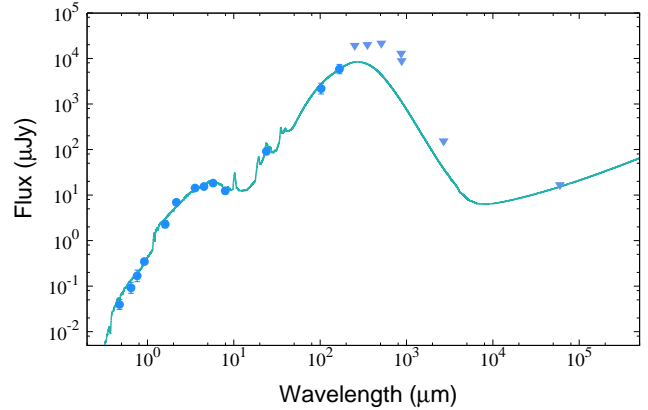,width=0.5\textwidth}
\end{center}
\vskip -4 mm
\caption[]{The best-fitting  Spectral Energy Distribution (SED) for the host galaxy of GRB 080207 obtained using \textit{LePhare}. The x-axis is wavelength in the observer frame. The circles present the photometry values of the GRB host while the triangles show the upper limit photometries.   
}
\label{fig:sed}
\end{figure}

\section{RESULTS}
\label{sec:res}

\subsection{Modeling the Spectral Energy Distribution of the host}
\label{sec:sed}

The host galaxy of GRB 080207 is a massive galaxy with a stellar mass  reported in a range of  $10^{11.05}-10^{11.17}\,\rm M_{\odot}$ 
\citep[][]{Hunt11-2011ApJ...736L..36H, Hunt14-2014A&A...565A.112H, Svensson12-2012MNRAS.421...25S, Perley13-2013ApJ...778..128P}. The reported star formation rate (SFR) of the galaxy varies in a large range of $46-416\,\rm M_{\odot}\,yr^{-1}$,  measured from different methods \citep[][]{Svensson12-2012MNRAS.421...25S, Kruhler12-2012ApJ...758...46K, Hunt14-2014A&A...565A.112H, Perley13-2013ApJ...778..128P}. Considering that this  galaxy is extremely red and dusty, its far-Infrared luminosity ($L_{\rm FIR}$) will provide the best estimate for its SFR.  

We add the  two upper limits from our observations with APEX and Plateau de Bure / NOEMA  to all available photometry   from UV up to radio  \citep[][]{Hunt11-2011ApJ...736L..36H, Perley13-2013ApJ...778..128P, Hunt14-2014A&A...565A.112H} and use   \textit{LePhare}   \citep[][]{1999MNRAS.310..540A} to model the Spectral Energy Distribution (SED) of the host galaxy. We use the radio band photometry  of the galaxy at 5.2 GHz from \citet[][]{Perley13-2013ApJ...778..128P} as an upper limit in our SED modeling  since the reported value could have  contamination from the radio-afterglow of the GRB itself \citetext{\citealp[see][for details on the radio photometry of the host]{Perley13-2013ApJ...778..128P}, \citealp[also see][for examples of the radio afterglow contamination several years after GRB events]{Perley17-2017MNRAS.465..970P}}.

\begin{table}
\begin{center}
\caption{
The properties of GRB 080207 host galaxy obtained from SED modeling. We use  the calibration from \citet[][]{Kennicutt12-2012ARA&A..50..531K} in order to obtain the $\rm SFR_{FIR}$ from $L_{\rm FIR}$. 
}
\label{tab:tab1}
\begin{tabular}{cccc}
\hline
$M_{*}$      & $E(B-V)$ & $\log (L_{\rm FIR}/\rm L_{\odot})$ & $\rm SFR_{FIR}$  \\
($\rm M_{\odot}$)                             &        &                             &($\rm M_{\odot}\, yr^{-1}$)  \\
\hline
&&&\\
($9.1^{+1.4}_{-2.2}) \times 10^{10}$ & 0.5 &12.18$^{+0.06}_{-0.10}$ & 260$^{+39}_{-53}$ \\  
&&&\\
\hline         		  
\end{tabular}
\end{center}
\end{table}

For modeling the stellar emission from the host galaxy, we  use  the  Stellar Population  Synthesis  templates developed by \citet{Bruzual03-2003MNRAS.344.1000B}. We limit our   models to the $\rm BC03_{-}m62$ library -- the templates with solar metallicity since the measured metallicity for GRB 080207 host galaxy  is   1.1 $\rm Z_{\odot}$ \citep[][]{Kruhler15-2015A&A...581A.125K}. We use the Chabrier IMF \citep[][]{2003PASP..115..763C} and consider both  the exponentially declining star formation history (SFR  $\propto {\rm e}^{-t/\tau}$) with $\tau$ values of 0.1, 0.3, 1, 2, 3, 5, 10, 15, and 30 Gyr, as well as the delayed star formation history (SFR $\propto -t\,{\rm e}^{t/\tau}$). The  dust extinction law from  \citet{Calzetti00-2000ApJ...533..682C} is applied to the templates with $E(B-V)$ values ranging between 0.0 and 1.2 with a step size of 0.1. Using \textit{LePhare} we separately model the SED of the galaxy in  infrared-radio wavelengths with templates from \citet[][]{Chary01-2001ApJ...556..562C}. 

The best-fitting SED model for the host galaxy is presented in Fig. \ref{fig:sed}. Estimated  galaxy   properties based on the SED modeling are provided in Table \ref{tab:tab1}. 
We use the calibration from \citet[][]{Kennicutt12-2012ARA&A..50..531K} and estimate  the $\rm SFR_{FIR}$ of the galaxy to be $260^{+39}_{-53}\,\rm M_{\odot}\,yr^{-1}$ from  the measured FIR luminosity of ${\rm\log}{(L_{\rm FIR}/\rm L_{\odot})} = {12.18}^{+0.06}_{-0.10}$. Our stellar mass measurement is consistent with the values obtained by previous studies \citep[][]{Hunt11-2011ApJ...736L..36H, Hunt14-2014A&A...565A.112H, Svensson12-2012MNRAS.421...25S, Perley13-2013ApJ...778..128P}. 
The host galaxy of GRB 080207 happens to be  a main-sequence star-forming galaxy \citep[][]{Elbaz11-2011A&A...533A.119E}. It also follows  the mass-metallicity relation of the general star-forming galaxy population with similar redshifts \citep[see][]{Arabsalmani18-2018MNRAS.473.3312A}.

\subsection{Molecular gas properties of the host}
\label{sec:mol}

\begin{figure}
\begin{center} \hskip -5mm
\psfig{file=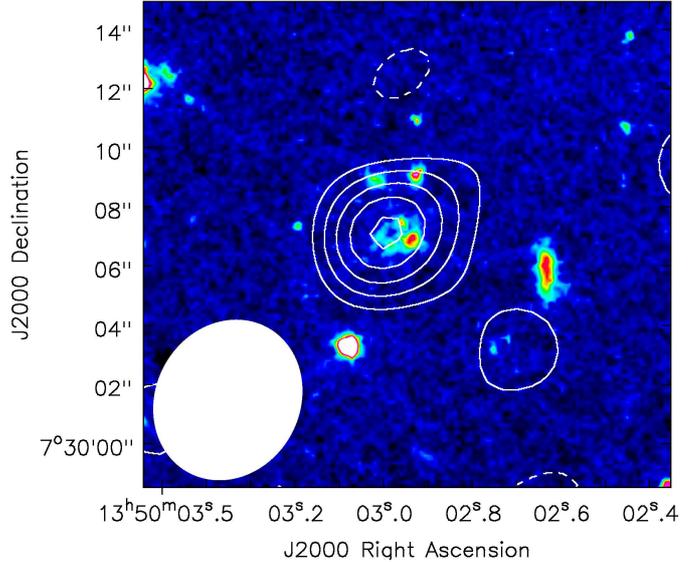,width=0.5\textwidth}
\end{center}
\vskip -4 mm
\caption[]{The velocity-integrated CO(3-2) map  (in contours) overlaid on the \textit{HST} image of the host galaxy. The beam-size is $4.8\arcsec \times 5.5\arcsec$. The outermost contour is at  $2\sigma$ significance, with subsequent contours  in steps of $1\sigma$. The negative contours  are at $-2\sigma$ level. 
}
\label{fig:co-map}
\end{figure}

We detect the CO(3-2) emission line from the host galaxy of GRB 080207 with a $7\sigma$ significance (all the measurements for the molecular gas properties of the GRB host are listed in  Table \ref{tab:tab}).  
Fig. \ref{fig:co-map} shows  the velocity-integrated CO(3-2)  map (in contours) overlaid on the \textit{Hubble Space Telescope} (\textit{HST})  image of the host galaxy  (with the WFC3/F110W filter, centred at 11534.459 \AA).

\begin{figure}
\begin{center} \hskip -5mm
\psfig{file=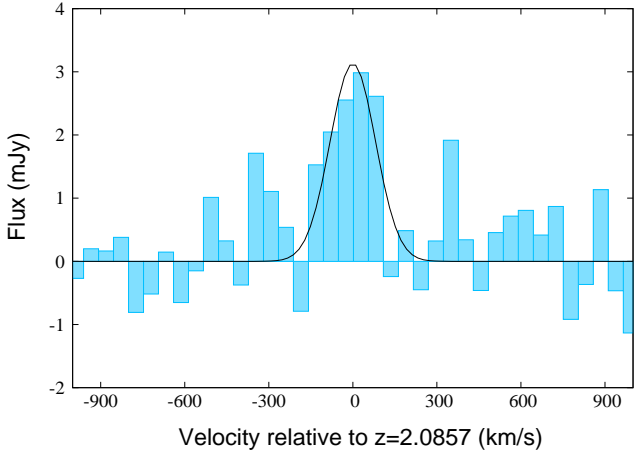,width=0.5\textwidth}
\end{center}
\vskip -4 mm
\caption[spectrum]{The CO(3-2)  emission line from the host galaxy of GRB 080207 detected by Plateau de Bure / NOEMA with a channel-width of 53 \kms.  The best-fitting  Gaussian to the line is shown with a black line. 
}
\label{fig:spec}
\end{figure}

As the source is spatially unresolved, we use the fluxes from the beam centred  on the source position only.  
The CO(3-2) line flux  in the velocity space (relative to $z$ = 2.0857) is shown in Fig. \ref{fig:spec}.  From the Gaussian fit to the CO line, we measure a redshift of  $z$ = $2.0857 \pm 0.0002$, consistent with the redshift of $z$ = $2.0858 \pm 0.0003$ 
derived from the nebular emission lines in the host spectrum  by \citet{Kruhler12-2012ApJ...758...46K}. 
We measure a velocity-integrated flux density of $0.63\pm0.08$ Jy\,\kms\, over five channels   covering a velocity spread  of 267 \kms. 
This leads to  a  brightness temperature luminosity, as defined in \citet{Obreschkow09-2009ApJ...702.1321O},  of $L^{\rm T}_{\rm CO(3-2)}= (1.54\pm0.20) \times 10^{10}\, \rm K\,km\,s^{-1}\,pc^2$, where 
\begin{eqnarray}
\nonumber
L^{\rm T} {(\rm K\,km\,s^{-1}\,pc^2) = 3.255 \times 10^{7} \big (\nu_{line, obs} (\rm GHz)\big)^{-2}} \cr
\times \big(D_L {(\rm Mpc)} \big)^{2} (1+z)^{-3}\big(S\Delta v {(\rm Jy\,kms^{-1}}) \big),  
\end{eqnarray}
same as the $L'$ defined in \citet{Solomon07-1997ApJ...478..144S}. 
From the best-fitting Gaussian we obtain a full-width-at-half-maximum (FWHM) of  $191\pm 35$ \kms\, for the CO(3-2) line.

\begin{figure*}
\captionsetup[subfigure]{labelformat=empty}
\begin{center}$
\begin{array}{cc} \hskip 0mm
\subfloat[]
{\includegraphics[trim = 0mm 0mm 0mm 0mm, clip, width=0.57\textwidth, angle=0]{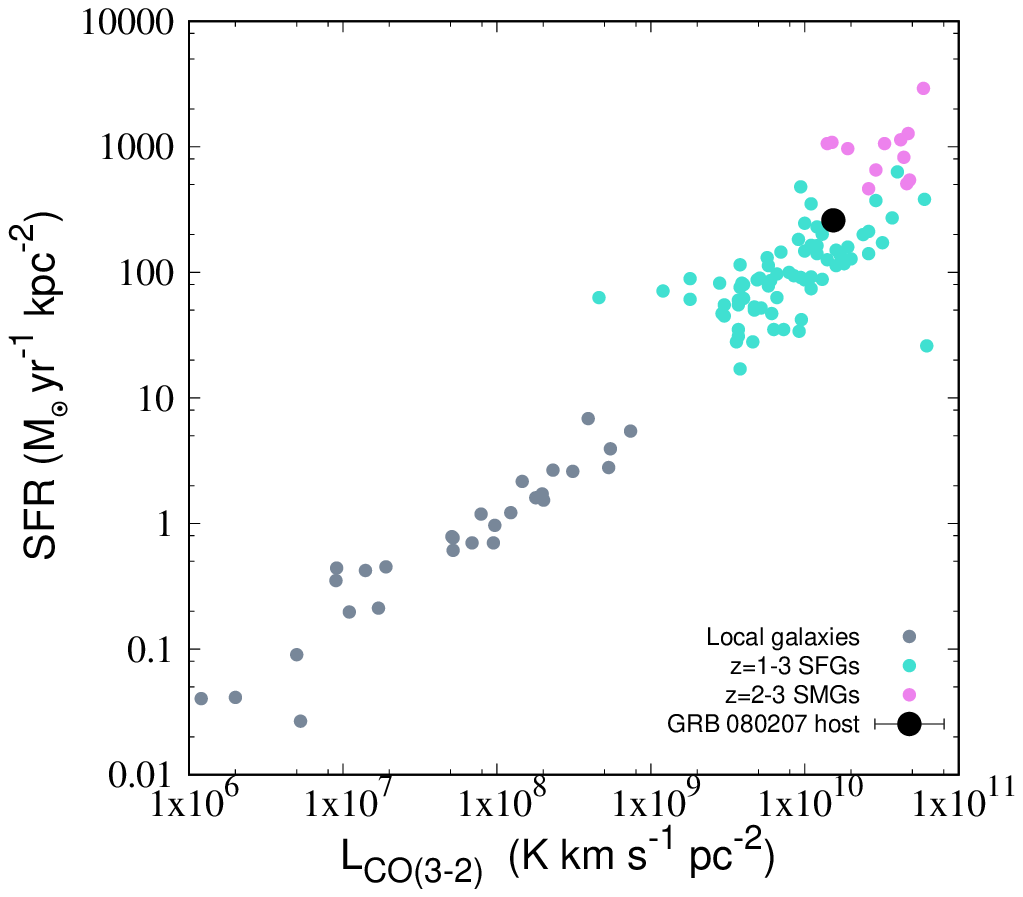}} 
& \hskip -17mm

\subfloat[]
{\includegraphics[trim = 0mm 0mm 0mm 0mm, clip, width=0.57\textwidth, angle=0]{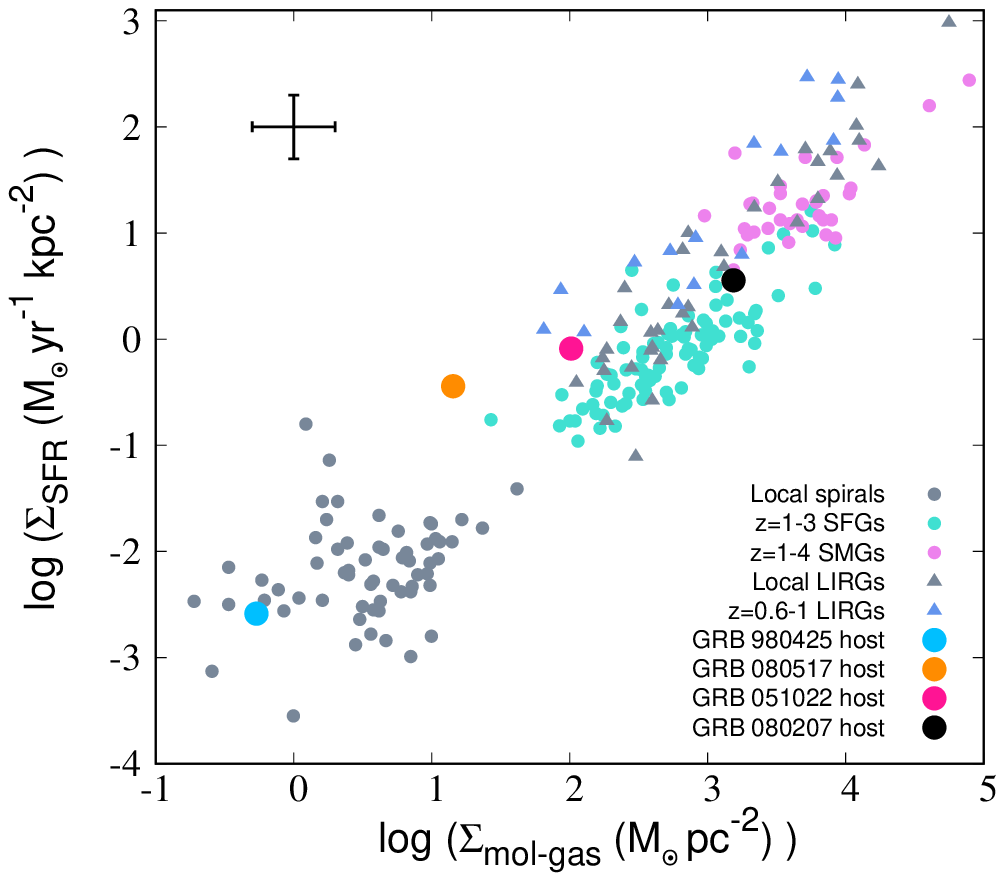}}
\end{array}$
\end{center}
\vskip -7.7 mm
\caption[]{
\textit{Left panel:} 
Observed 
$L^{\rm T}_{\rm CO(3-2)}$  versus SFR, for  local Star-Forming Galaxies \citep[SFGs, in gray, from][]{Wilson12-2012MNRAS.424.3050W}, $z = 1-3$ SFGs \citep[in green, from ][]{Tacconi10-2010Natur.463..781T, Tacconi13-2013ApJ...768...74T}, and $z = 2-3$ luminous Sub-Millimetre  Galaxies  \citep[SMGs, in violet, from][]{Bothwell13-2013MNRAS.429.3047B}. The $L^{\rm T}_{\rm CO(3-2)}$ for GRB 080207 host (this work) is marked with a black circle. 
\textit{Right panel:} 
Kennicutt-Schmidt relation,  for local spiral galaxies  \citep[gray circles, from][]{Kennicutt98-1998ApJ...498..541K}, local Luminous Infrared Galaxies \citep[LIRGs, gray triangles, from][]{Kennicutt98-1998ApJ...498..541K},   $z = 0.6-1$ LIRGs \citep[blue triangles, from][]{Combes13-2013A&A...550A..41C}, $z = 1-3$ SFGs \citep[cyan  circles, from ][]{Daddi10-2010ApJ...713..686D, Genzel10-2010MNRAS.407.2091G, Tacconi10-2010Natur.463..781T, Tacconi13-2013ApJ...768...74T}, and $z = 1-4$ SMGs \citep[violet circles, from][]{Genzel10-2010MNRAS.407.2091G, Bothwell13-2013MNRAS.429.3047B}. 
Note that for all the sample galaxies the molecular gas mass includes both the H$_2$ and He mass (with $M_{\rm He} = 0.36 M_{\rm H_2}$) and also $\alpha_{\rm CO}=4.36$ $\rm M_{\odot}(\rm K\,km\,s^{-1}\,pc^2)^{-1}$ is assumed.  Also, for  galaxies taken from  \citet[][]{Bothwell13-2013MNRAS.429.3047B} a typical half-light-radius  of 3.0 kpc is assumed in order to estimate the surface densities. 
The error-bar on the top left  represents a typical error on the data points. 
The large  black circle represents the host galaxy of GRB 080207 at $z$ = 2.086 and the three other GRB hosts with detected 
CO emission lines are marked with orange, red, and  blue large circles,  assuming  $\alpha_{\rm CO} = 4.36$ $\rm M_{\odot}(\rm K\,km\,s^{-1}\,pc^2)^{-1}$ for all four host galaxies. Note that this assumption is likely to result in underestimating the molecular gas masses of  the GRB hosts with sub-solar metallicities and consequently providing lower limits for the surface-densities of their molecular gas mass  (see Section \ref{sec:dis} for details).  

}
\label{fig:KS}
\end{figure*}

The left panel of Fig. \ref{fig:KS} shows the  SFR-$L^{\rm T}_{\rm CO(3-2)}$ correlation for star-forming galaxies (SFGs) with the position of GRB 080207 host galaxy marked 
with a black circle. Our target host clearly falls on the same region as typical SFGs  with similar redshifts. 
Although the SFR of our target galaxy  happens to be on the higher side given its stellar mass, it places the galaxy   amongst  the 
main-sequence star-forming galaxies \citep[see Fig.1 of][]{Rodighiero11-2011ApJ...739L..40R}.  
Therefore,   in order to convert the  measured $L^{\rm T}_{\rm CO(3-2)}$ to a brightness temperature luminosity  of CO(1-0) we use a $R_{31} = L^{\rm T}_{\rm CO(3-2)} / L^{\rm T}_{\rm CO(1-0)}$ 
of 0.6 from \citet[][]{Carilli13-2013ARA&A..51..105C} which is the value for main-sequence star-forming galaxies \citep[see also  
][]{Dannerbauer09-2009ApJ...698L.178D, Daddi15-2015A&A...577A..46D}   
and obtain   $L^{\rm T}_{\rm CO(1-0)} = 2.56 \times 10^{10}\, \rm K\,km\,s^{-1}\,pc^2$. 

To estimate the  molecular gas mass for the host galaxy we need to assume a reasonable CO-to-molecular-gas 
conversion factor ($\alpha_{\rm CO}$). The GRB host has a 12+log[O/H] of 8.7 \citep[][]{Kruhler15-2015A&A...581A.125K}, equivalent to a metallicity of 1.1 solar. 
Using the Galactic  $\alpha_{\rm CO}$ of 4.36 $\rm M_{\odot}(\rm K\,km\,s^{-1}\,pc^2)^{-1}$, 
we estimate   a molecular gas mass of $1.1 \times 10^{11}\,\rm M_{\odot}$. With a stellar mass of $9.1 \times 10^{10}\, \rm M_{\odot}$ (see Table. \ref{tab:tab1}), we find the molecular gas fraction 
of the host to be $f_{\rm mol-gas} \sim 0.5$, where   $f_{\rm mol-gas}$ is $M_{\rm mol-gas} / (M_{\rm mol-gas} + M_*)$. This molecular gas fraction   is typical of SFGs with similar stellar masses and redshifts 
\citep[see Fig.6 of][for the gas fractions of  star forming galaxies at $z = 1-3$]{Tacconi13-2013ApJ...768...74T}. 
We also find  a molecular-gas-depletion timescale ($M_{\rm mol}/\rm SFR$) of 0.43 Gyr for  the GRB host which is at the peak 
of the depletion timescale distribution of    SFGs at similar redshifts \citep[see Fig. 7 of ][]
{Tacconi13-2013ApJ...768...74T}. 

We use the half-light-radius of $R_{0.5}=3.4$ kpc  \citep[from][remeasured and confirmed by us]{Svensson12-2012MNRAS.421...25S} and  
calculate a
SFR  surface-density of $\Sigma_{\rm SFR} = 3.6\,\rm M_{\odot}\rm  yr^{-1} kpc^{-2}$ and a  molecular gas surface-density of  $\Sigma_{\rm mol-gas} = 1.5 \times 10^{3}\,\rm M_{\odot} pc^{-2}$ for the host. 
The right panel  of Fig. \ref{fig:KS} shows the  Kennicutt-Schmidt (K-S) relation \citep[][]{Kennicutt98-1998ApJ...498..541K}    
with a black circle representing the GRB host. The host galaxy of GRB 080207  clearly follows  the K-S relation of the SFGs. 


\begin{table*}
\begin{center}
\caption{
The molecular gas properties  of GRB 080207 host galaxy.
}
\label{tab:tab}
\begin{tabular}{ccccccc}
\hline
Redshift  & $\rm FWHM_{CO(3-2)}$ &  $(S \Delta v) _{\rm CO(3-2)}$ & $L^T_{\rm CO(3-2)}$ & $M_{\rm mol-gas}$ & $\Sigma_{\rm SFR}$ & $\Sigma_{\rm mol-gas}$  \\
  & (\kms) &  (Jy \kms) & (K \kms pc$^2$) & ($\rm M_{\odot}$) & ($\rm M_{\odot}\,\rm yr^{-1}$\,kpc$^{-2}$) & ($\rm M_{\odot}$\,pc$^{-2}$) \\
\hline
&&&&&& \\
2.086 & $191 \pm 35$  & $0.63\pm0.08$ & $(1.54\pm0.20) \times 10^{10}$ & $1.1 \times 10^{11}$  & 3.6 &  $1.5 \times 10^3$  \\ 
&&&&&&\\ 
\hline         		  
\end{tabular}
\end{center}
\end{table*}

\subsection{Kinematics  of molecular versus ionized gas}
\begin{figure}
\begin{center} \hskip -7.8mm
\psfig{file=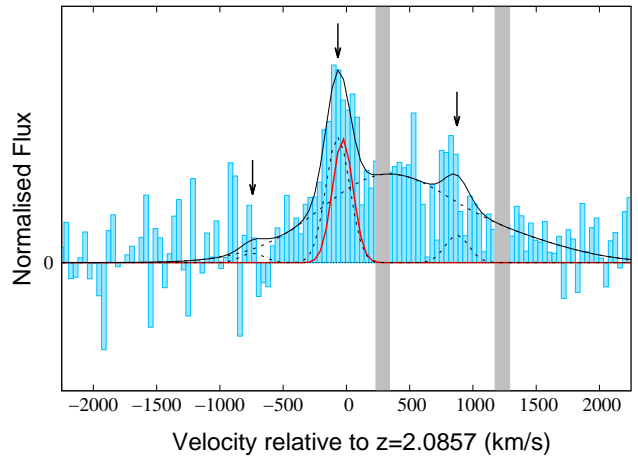,width=0.52\textwidth}
\end{center}
\vskip -4 mm
\caption[spectrum]{
The \ha and N{\sc II} emission lines from the host galaxy of GRB 080207.  
 The best-fitting   2-component model is shown with the black-solid line.   The narrow and broad  components are marked with 
black-dotted lines. The best-fitting Gaussian  to the CO(3-2) emission 
line is shown with the solid red line, scaled in y-axis such that its peak matches that of the narrow component of the \ha line.  The vertical gray regions    present the wavelengths with contamination from the sky-lines. The three arrows mark the peaks of the narrow  \ha, [N{\sc ii}]6548, and [N{\sc ii}]6584   lines.  
}
\label{fig:ha}
\end{figure}

In this section we  compare the kinematics of the molecular gas with that of the ionized gas through a comparison between the structure and velocity widths of CO(3-2) and the bright nebular emission lines from the galaxy. For this  we use the 1-D extracted  VLT/X-shooter spectrum of the host galaxy presented in \citet[][]{Kruhler15-2015A&A...581A.125K} and available at  VizieR database \citep[see][for the details  of the data reduction]{Kruhler12-2012ApJ...758...46K, Kruhler15-2015A&A...581A.125K}.  A part of the NIR spectrum, covering the wavelengths of  \ha and [N{\sc ii}]6548,6584 emission lines,   is shown in Fig. \ref{fig:ha}. 
The \ha emission line  seems to have an extension in the red side, spreading over several hundreds of \kms. This  is clearly apparent in the 2-D spectrum of the \ha line presented in Fig. 3 of  \citet[][]{Kruhler12-2012ApJ...758...46K}. Although the overall S/N ratio of the X-shooter NIR  spectrum of the host is not  high, the extension spreading over the velocity range between $350$ and $600$ \kms \, relative to the redshift of the host, is detected with 10$\sigma$ significance.   
While the CO(3-2) emission line has a FWHM of $191\pm 35$ \kms, a single Gaussian fit to the \ha line results in a FWHM of 790$\pm$60 \kms.  This is an evidence for the significant contribution from a broad emission in \ha   which is not present in the CO(3-2) emission line.


 
To investigate this further, we  model the galaxy spectrum (shown in Fig. \ref{fig:ha})    by a two component system. We simultaneously fit  double-Gaussian functions (summation of two independent Gaussians)  to each of the \ha and [N{\sc ii}]6548,6584 lines. We allow the velocity widths of the two components to vary, but for each component set the line-broadening of the \ha and [N{\sc ii}] lines to be the same. The peak fluxes of the \ha and [N{\sc ii}]6584 lined associated with the  two components   are free to vary, but  we fix the  [N{\sc ii}]6584/[N{\sc ii}]6564 ratio to the theoretical value of 2.92  \citep[][]{Acker98-1989Msngr..58...44A, Storey00-2000MNRAS.312..813S} for both components. We therefore have eight free parameters: the \ha line centres and  broadenings  associated with the two components (four parameters),  and the peak fluxes of the \ha and [N{\sc ii}]6584 lines for the two components (four parameters). 
  
We find the best-fitting  model (with a reduced $\chi^2$ of 1.2) to be the combination of a narrow and a broad component. The narrow component (detected at $\sim 5\sigma$ significance) is centred  at a velocity of $-67 \pm 13$ \kms\, relative to the redshift obtained form the CO(3-2) emission line ($z=2.0857$), or equivalently, is centred  at the redshift of 2.0850$\pm$0.0001. It has  a  FWHM of  207$\pm$35 \kms\, \citep[corrected for the typical spectral resolution  of 50 \kms\, in the NIR arm of X-shooter, see][]{Arabsalmani15-2015MNRAS.446..990A},  consistent  with the FWHM   of the CO(3-2) emission line. 
As shown in Fig. \ref{fig:ha}, the narrow component  very well matches the  CO(3-2) emission  line in velocity space. Note that the peak flux of the CO(3-2) line in Fig. \ref{fig:ha} is scaled to match the peak flux of the narrow component of the \ha emission. 

The second component, centred at $320 \pm 110$ \kms\, relative to $z=2.0857$, and detected at $\sim 3\sigma$ significance,  is quite    broad  with a FWHM of $1160 \pm 330$\kms. The large FWHM of this component is consistent with the signature of a powerful outflow \citep[see][]{Newman12-2012ApJ...752..111N, Newman12-2012ApJ...761...43N}. Note that outflows are not necessarily symmetric and depending on their orientation  they can be detected  as red-shifted or blue-shifted components \citep[see][for an example of red-shifted outflowing gas]{Feruglio15-2015A&A...583A..99F}. The broad component accounts for  $77\% \pm 30\%$ of the total \ha flux. This is similar to the case of an individual clump in the $z \sim 2$ clumpy SFG presented in \citet[][]{Newman12-2012ApJ...752..111N}, where the \ha emission from the clump is dominated by a broad component associated  with a strong  outflowing  gas.

For both  components  the [N{\sc ii}]6584/\ha ratio is roughly 0.2-0.3, which is typical for star-formation-driven outflows, and less than the ratios  expected for Active galactic nucleus (AGN) driven outflows  
\citep[][]{Newman12-2012ApJ...752..111N, Newman12-2012ApJ...761...43N, ForsterSchreiber14-2014ApJ...787...38F}. In addition,  the large offset between the two components  is against the presence of an AGN being responsible for the broad emission  \citep[see][]{Genzel14-2014ApJ...796....7G}. The low S/N of the NIR spectrum does not allow  performing a detailed analysis for the H$\beta$ and oxygen  lines in order to further investigate the possible contribution from an AGN. We should mention that our SED modeling presented in section \ref{sec:sed} does not favor the presence of an AGN. Also note that GRB host galaxies typically have high SFR surface-densities \citep[][]{Kelly14-2014ApJ...789...23K} which predicts  strong star-formation driven outflows  from their compact star-forming regions \citep[see][]{Lagos13-2013MNRAS.436.1787L, Arabsalmani18-2018MNRAS.473.3312A}.

The presence of a broad component in  the \ha emission line is not uncommon in SFGs  at $z \sim 2$ \citep[][]{Newman12-2012ApJ...752..111N, Newman12-2012ApJ...761...43N, ForsterSchreiber14-2014ApJ...787...38F, Genzel14-2014ApJ...796....7G}. It is notable though that in the SFG sample of  \citet[][]{Genzel14-2014ApJ...796....7G} the average flux ratio of the broad  to narrow component is $\sim 0.4$ while for the host of GRB 080207 this ratio is $\sim 3.4$. This  is also larger than the broad to narrow flux ratios in all the SFGs presented in  \citet[][see their Fig.2]{Newman12-2012ApJ...761...43N}. The characteristics of the \ha emission line from the host galaxy of GRB 080207 is more similar to those of the mentioned  individual clump  in the $z\sim2$ clumpy SFG presented in \citet[][]{Newman12-2012ApJ...752..111N} where the presence of a super-wind  is proposed.

Given the  large separation of $\sim 390$ \kms\, between the narrow and the broad components in our fit,  it is also conceivable that the broad emission in \ha is associated with a companion galaxy  interacting with the GRB host. This is   consistent with the irregular morphology of the galaxy (Fig. \ref{fig:co-map}) which suggests  the possibility of  an interacting system \citep[][]{Svensson12-2012MNRAS.421...25S}. However, considering the  large velocity width of the broad  component, its associated gas  is unlikely to be   gravitationally bound and is  plausibly  originating from an outflowing gas either in the GRB host or in an interacting companion. Spatially resolved follow up observations  are required  in order to find out  the nature of this broad emission.

\section{Discussion on the molecular gas content of GRB hosts}
\label{sec:dis}

It has been suggested that GRB host galaxies  are deficient    in  molecular gas. This is mainly based on the measured molecular gas content of  three GRB host galaxies: GRB 980425 at $z$ = 0.0087, GRB 080517 at $z$ = 0.089, and GRB 051022  at $z$ = 0.81. 

For the host galaxy of GRB 980425,   \citet{Michalowski16-2016A&A...595A..72M} report the molecular gas mass to be a factor of 3-7 less than expected  from the  atomic gas content  and stellar mass  of the galaxy. However, their comparison relies on the  empirical relations of large spiral  galaxies with measured atomic and molecular gas masses. This is while GRB 980425 host has a stellar mass of $M_* = \rm 10 ^{8.6} M_{\odot}$ \citep[][]{Michalowski14-2014A&A...562A..70M} which is typical of  dwarf galaxies. 
If we compare its molecular gas mass  with those of local dwarf galaxies with similar stellar masses, we find that the host galaxy is quite normal in its molecular gas content \citep[see Fig. 5 in][and note  that  the $M_{\rm mol-gas}/M_{*}$  ratio  flattens at low  stellar masses]{Grossi16-2016A&A...590A..27G}.   This GRB host  clearly   follows  the $L_{\rm CO(2-1)}-\rm SFR$ relation. 
Using its  half-light radius $R_{0.5} = 4$ kpc \citep[][]{Michalowski14-2014A&A...562A..70M} and  assuming a Galactic $\alpha_{\rm CO}$, it also appears  to follow the K-S relation   (see the right panel  of Fig. \ref{fig:KS}). Note that  $\alpha_{\rm CO} \gtrsim 10.0$ \citep[][]{Schruba12-2012AJ....143..138S, Genzel12-2012ApJ...746...69G} is a  more appropriate assumption for  this galaxy given its low metallicity of $Z= 0.3 \rm Z_{\odot}$ \citep[][]{Christensen08-2008A&A...490...45C}. This  will shift the position of this host towards even larger molecular gas surface densities in the right panel of Fig. \ref{fig:KS}. We should mention that  this GRB host  also obeys  the  $\Sigma_{\rm total-gas}-\Sigma_{\rm SFR}$ \citep[][]{Kennicutt98-1998ApJ...498..541K} when considering its atomic+molecular gas mass \citep[][]{Arabsalmani15-2015MNRAS.454L..51A, Michalowski15-2015A&A...582A..78M}. 

A similar case is applicable to the host galaxy of  GRB 051022  with a metallicity of $Z= 0.6 \rm Z_{\odot}$. This host \citep{Hatsukade14-2014Natur.510..247H} is shown in the $\Sigma_{\rm mol-gas}-\Sigma_{SFR}$ plane in the right panel of Fig. \ref{fig:KS} with a Galactic $\alpha_{\rm CO}$ and  assuming $R_{0.5} = 2.5$  kpc 
\citep[based on the relation between $R_{0.5}$ and stellar mass from][]{Shibuya15-2015ApJS..219...15S}. A metallicity-dependent $\alpha_{\rm CO}$ for this galaxy  will shift its position towards  larger molecular gas surface densities on the K-S plane. It will also  lead to a   molecular gas fraction  consistent with those of SFGs with similar specific SFR values and redshifts \citetext{\citealp[see Fig.6 of][]{Tacconi13-2013ApJ...768...74T}}. 

For the host galaxy of  GRB 080517, \citet[][]{Stanway15-2015ApJ...798L...7S} assumed a Galactic $\alpha_{\rm CO}$  and found the galaxy to have a  molecular gas fraction  similar to those of local SFGs,  but with shorter molecular gas depletion time compared to the average value for the local SFGs with similar stellar masses. This host too is shown on the right panel of Fig. \ref{fig:KS}  on the $\Sigma_{\rm mol-gas}-\Sigma_{\rm SFR}$ plane with $R_{0.5} = 2.7$ taken from \citet{Stanway15-2015MNRAS.446.3911S}. 
Note that there is no metallicity measurement available for this host. Given its stellar mass of $M_* = 10^{9.6}\, \rm M_{\odot}$ \citep[][]{Stanway15-2015MNRAS.446.3911S} this host is expected to have a sub-solar metallicity \citep[see][for the mass-metallicity relation of GRB host galaxies]{Arabsalmani18-2018MNRAS.473.3312A} and hence assuming a Galactic $\alpha_{\rm CO}$  should provide a lower limit for its molecular gas mass.

We emphasize  that when investigating the issue of molecular gas deficiency  one should  be cautious  of the typical low metallicities of GRB hosts and hence   take into account the  correct assumptions about the CO-to-molecular-gas conversion factor. In addition, any comparison should be done with similar galaxies in terms of overall properties. 

For  GRB 080207 host galaxy, as discussed in Section \ref{sec:mol},  its  molecular gas properties are typical of SFGs with similar stellar masses and redshifts. 
Increasing the sample size of GRB host galaxies with measured molecular gas mass is required in order to draw any  conclusion on the content of molecular gas in GRB hosts compared  to the general SFG population. 


\section{Summary}
\label{sec:sum}
In this paper we present the molecular gas properties of   GRB 080207 host galaxy  at $z=2.086$. This is the fourth GRB host  with molecular gas detected in emission, and the first beyond redshift 1. We show  that the host galaxy of GRB 080207  has molecular gas properties similar to those 
of the  SFG population at similar redshifts. We also discuss  that the reported molecular 
gas deficiency in other three  GRB hosts  with detected CO emission lines may not be real.  
We emphasize  that when investigating the issue of molecular gas deficiency one should  be cautious  to compare   similar galaxies in terms of overall properties, and to take into account the  correct assumptions 
about the CO-to-molecular-gas conversion factor  for the galaxy in question. 

We also compare the kinematics of the CO(3-2) emission line with that of the \ha emission line from  the host galaxy of GRB 080207. We find the \ha emission to have contribution from two separate components, a narrow and a broad one. The narrow component appears to match well with the CO(3-2) emission line in velocity space. We speculate that the broad component is associated with a strong outflowing gas in the GRB host or in a companion galaxy interacting with the host.

\section*{Acknowledgments}

We would like to thank Diego Gotz and Giulia Migliori  for their  support and help, and Attila Kovacs for valuable help with CRUSH software  in data reduction. 
M. A. would like to thank Sambit Roychowdhury, Bernd Husemann, Daniel Perley, and Johan Fynbo  for valuable discussions. 
We acknowledge the financial support from:  UnivEarthS Labex program at Sorbonne Paris 
Cit\'e (ANR-10-LABX-0023 and ANR-11-IDEX-0005-02) for M.A., French National Research Agency (ANR) under contract ANR-16-CE31-0003 BEaPro for 
S.D.V.,  the 
MINECO
under the 2014 Ram\'on y Cajal program MINECO RYC-2014-15686 for H.D., NOVA and 
NWO-FAPESP grant for advanced instrumentation in astronomy for J.J., the European Union Horizon 2020 research and innovation programme under the Marie Sklodowska-Curie grant agreement No 664931 for C.F..  This work is based on observations carried out under project number w0c9/2013 with the IRAM NOEMA Interferometer. IRAM is supported by INSU/CNRS (France), MPG (Germany) and IGN (Spain).

\bibliographystyle{mnras}
\bibliography{adssample}

\end{document}